\documentclass[preprint,eqsecnum,preprintnumbers,nofootinbib]{revtex4}
\usepackage{amsmath,amssymb,amsfonts,times,graphicx,color,verbatim}
\pdfoutput=1

\newcommand{\beq}{\begin{equation}}
\newcommand{\eeq}{\end{equation}}
\def\bea{\begin{eqnarray}}
\def\eea{\end{eqnarray}}

\begin{document}

\begin{titlepage}
\vspace*{20mm}
\begin{center}
{\Large \bf Universality of Gravity from Entanglement}

\vspace*{15mm}
\vspace*{1mm}
Brian Swingle${}^a$ and Mark Van Raamsdonk${}^b$

\vspace*{1cm}
{\it ${}^a$
{Department of Physics, Harvard University, Cambridge MA 02138, USA}

 ${}^b$
{Department of Physics and Astronomy,University of British Columbia\\
6224 Agricultural Road,Vancouver, B.C., V6T 1W9, Canada} \\ }

\vspace*{1cm}

{\bf Abstract}
\end{center}
The entanglement ``first law'' in conformal field theories relates the entanglement entropy for a ball-shaped region to an integral over the same region involving the expectation value of the CFT stress-energy tensor, for infinitesimal perturbations to the CFT vacuum state. In recent work, this was exploited at leading order in $N$ in the context of large $N$ holographic CFTs to show that any geometry dual to a perturbed CFT state must satisfy Einstein's equations linearized about pure AdS. In this note, we investigate the implications of the leading $1/N$ correction to the exact CFT result. We show that these corrections give rise to the source term for the gravitational equations: for semiclassical bulk states, the expectation value of the bulk stress-energy tensor appears as a source in the linearized equations. In particular, the CFT first law leads to Newton's Law of gravitation and the fact that all sources of stress-energy source the gravitational field. In our derivation, this universality of gravity comes directly from the universality of entanglement (the fact that all degrees of freedom in a subsystem contribute to entanglement entropy).

\end{titlepage}

\section{Overview}

Holographic duality posits that certain conformal field theories (CFTs) are exactly equivalent to various higher-dimensional theories of quantum gravity, with a precise correspondence between states, observables, and dynamics.\footnote{This duality has now been extended far beyond the class of CFTs and may even include all field theories provided we define ``quantum gravity" broadly enough.} While neither side is necessarily more fundamental than the other, we understand much better how to construct CFTs from microscopic regulated degrees of freedom. Thus, it is natural to ask how the dynamical higher dimensional spacetime emerges from the CFT physics.  A crucial idea that has appeared in recent years is that this emergence is deeply connected with quantum entanglement between degrees of freedom in the CFT \cite{swingle, mav1}.\footnote{For recent work in this direction, see \cite{arch,Balasubramanian:2013lsa,Maldacena:2013xja}.} In this note, we pursue this idea and build on recent progress to demonstrate that in a certain semi-classical limit the bulk gravitational metric obeys the linearized (about anti de Sitter space) Einstein equations \textit{universally coupled} to matter.\footnote{We focus here on the simplest theories for which the holographic entanglement functional is area.}  Here matter includes all bulk degrees of freedom (including perturbative gravitons) treated as quantum fields defined on a semi-classical background.  In fact, since the graviton stress tensor is quadratic in the perturbation from AdS, our linearized equation in some sense knows about the first non-linear correction as well.  Furthermore, we argue that assuming the bulk physics follows from a local Lagrangian and given our entanglement assumptions, the bulk action must be the non-linear Einstein action plus matter.

Our argument proceeds from the following basic result. Given any (d+1)-dimensional CFT, we can consider the reduced density matrix $\rho_B$ of a $d$-ball $B$ of radius $R$ and define the entanglement entropy associated to this region as the von Neumann entropy $S_B = - \text{tr}(\rho_B \ln{(\rho_B)})$. Starting from the CFT vacuum state and shifting to a nearly parallel state in the Hilbert space, it is straightforward to show \cite{chm, relative} that the variation in $\delta S_B$ in this entanglement entropy can be reexpressed as the variation in a certain energy $E_B$ (defined below) associated with the region $B$. Thus, we obtain an exact relation $\delta S_B = \delta E_B$ closely connected with the First Law of Thermodynamics.\footnote{See \cite{therm} for some earlier attempts to make such a connection.}

In \cite{eom,Faulkner:2013ica}, this relation was considered in the context of holographic conformal field theories, for which $\delta S_B$ and $\delta E_B$ may be translated via the holographic dictionary \cite{rt1,rt2,timedep} to geometrical observables in the dual gravity theory. Working at leading order in large $N$ (or equivalently in the limit $G_N \to 0$ in the gravitational theory), it was demonstrated that the CFT relation $\delta S_B = \delta E_B$ translates to a constraint on the geometries dual to near-vacuum CFT states that is precisely equivalent to the gravitational equations linearized about AdS. Thus, by assuming the holographic dictionary for entanglement entropy, it is possible to derive Einstein's equations at linear order from the physics of entanglement in the CFT.

Since the CFT relation $\delta S_B = \delta E_B$ holds exactly (i.e. not just at leading order in large N), it is interesting to understand the implications in the gravity theory of the subleading $1/N$ corrections. In fact, there is an enormous motivation do so: according to the usual holographic dictionary, the $1/N$ expansion in the CFT corresponds to the expansion on the gravity side in powers of $G_N$. The source term $8 \pi G_N T_{\mu \nu}$ on the right side of Einstein's equations appears at order $G_N$; thus, exploiting $\delta S_B = \delta E_B$ at the first subleading order in $1/N$, we should be able to derive the source term in the gravitational equations, i.e. the universal coupling of the stress-energy of the bulk fields to the metric. This is precisely what we find: including the $1/N$ corrections, the CFT relation $\delta S_B = \delta E_B$ corresponds directly to the linearized gravitational equations including the usual source term. Assuming that this source term remains local away from the limit of linearized perturbations to the state of the bulk matter fields, we recover all of Newtonian gravity directly from this simple constraint on entanglement entropies.

\subsection*{Holographic entanglement at subleading order}

The crucial ingredient for translating $\delta S_B = \delta E_B$ to a statement in the gravity theory including subleading terms in the $1/N$ expansion is the holographic dictionary for the entanglement entropy beyond the leading order.

To describe this formula, we need to specify which class of CFT states we are considering. Assuming our CFT is a large $N$ theory, there should be a subset of states $|\psi\rangle_{\text{CFT}}$ whose dual gravity interpretation is ``semi-classical''. That is, we can associate to $|\psi\rangle_{\text{CFT}}$ a pair $(M,|\psi\rangle_{\text{bulk}})$ consisting of a classical geometry $M$ and a bulk QFT state $|\psi\rangle_{\text{bulk}}$ living on this geometry, where $|\psi\rangle_{\text{bulk}}$ describes quantum fluctuations of all fields $\phi$ living in the bulk, including the metric perturbation.

Assuming such a state, our fundamental assumption (based on the proposal of Ryu and Takayanagi \cite{rt1,rt2,timedep,aitor} and the proposal of Faulkner, Lewkowicz, and Maldacena for the subleading correction \cite{Faulkner:2013ana}) is that the CFT entanglement entropy $S_A$ ($A$ is any region in the CFT) for the state $|\psi\rangle_{\text{CFT}}$ corresponds on the gravity side at leading and subleading order in $G_N$ to the expression
\beq \label{holoee}
S_A(|\psi\rangle_{\text{CFT}}) = \frac{|\tilde{A}|_M}{4 G_N} + S_\Sigma(|\psi\rangle_{\text{bulk}}),
\eeq
where $|\tilde{A}|_M$ is the area\footnote{More generally, we can have some Wald-like functional; for such theories, the gravitational equations that emerge are some more general covariant equations, e.g. involving higher powers of curvatures \cite{Faulkner:2013ica}.} of the bulk extremal surface $\tilde{A}$ with $\partial \tilde{A} = \partial{A}$ and $S_\Sigma(|\psi\rangle_{\text{bulk}})$ is the conventional entanglement entropy of bulk fields $\phi$ in their state $|\psi\rangle_{\text{bulk}}$, for the spatial region $\Sigma$ bounded by $A$ and $\tilde{A}$.\footnote{There are various choices for this spatial region $\Sigma$, corresponding to different time-slicings of the bulk spacetime. However, since these surfaces all have the same domain of dependence (causal development region), the fields restricted to any two choices $\Sigma$ and $\Sigma'$ represent the same degrees of freedom. Therefore the entanglement entropy $S_\Sigma$ is independent of the choice of surface.}

Starting with this corrected formula for holographic entanglement entropy, our analysis proceeds in much the same way as in \cite{eom,Faulkner:2013ica}. We start with a class of deformations of the CFT ground state which can be represented as a small deformation of the classical bulk geometry and a small variation of the bulk state of the $\phi$ fields.  We then compute both sides of the entanglement ``first law" $\delta S_R = \delta E_R$ using our main assumption (\ref{holoee}). The result is a nonlocal constraint involving the bulk metric and the variation of the bulk entanglement entropy for the region $\Sigma$. Fortunately, since $\Sigma$ corresponds (for some choice of coordinates) to the half-space in AdS, it is possible to prove a {\it bulk relation} $\delta S_\Sigma = \delta E_\Sigma$ relating the variation of the bulk entanglement entropy to an integral involving the bulk stress tensor. Using this, we can rewrite the nonlocal constraint coming from the CFT first law as a local bulk constraint, which is exactly the linearized Einstein equations with the bulk stress tensor as a source.

\subsection*{From linear to nonlinear}

Assuming that the local dynamical equations that we derive at the linearized level extend to some local dynamical equations that constrain the metric for larger deviations from pure AdS, it is almost immediate that these non-linear equations must be the full Einstein equations. Once we couple the linearized metric fluctuations to dynamical matter, then the full non-linear equations follow from demanding self-consistency of the resulting equations with conservation of energy and momentum.  There is a subtlety here, in that many non-linear actions may give rise to the same linearized two-derivative theory, but we need only appeal to our original assumption about the encoding of entanglement to uniquely fix the non-linear action.

\subsection*{Comments on the main assumption}

Eq. (\ref{holoee}) is a strong assumption, but not an unreasonable one.  The leading area term is ultimately the main assumption, and we will discuss partial success towards deriving it from more basic considerations later in the note.  However, once we have the bulk surface $\tilde{A}$, which for the balls we consider turns out to be a horizon (technically a bifurcation surface of a Killing horizon), then the appearance of $S_\Sigma(|\psi\rangle_{\text{bulk}})$ is very natural.  Indeed, there is a natural class of CFT observers associated with $A$ and a corresponding class of bulk observers associated with $\Sigma$ for which the horizon $\tilde{A}$ represents the boundary of their knowledge of the state of the $\phi$ fields.  Hence it is reasonable to suppose that in addition to the classical geometrical entropy given by the minimal surface area, we must also include the entropy due to these observers' ignorance of the physics behind the horizon.  We will have a few more comments concerning our main assumption at the end of the note.

\section{Basic machinery}

To begin, we recall the basic CFT result that we will use to derive the bulk equations, and discuss the holographic formula for entanglement entropy that allows us to translate the CFT result to the dual gravitational theory.

\subsection{The entanglement ``first law" for CFTs}

Our starting point is a general formula for the variation of entanglement entropy for a subsystem $A$ under an infinitesimal variation in the global state,\footnote{Specifically, we are considering states of the form $|\Psi_0 \rangle + \epsilon | \Psi_1 \rangle$ in the limit $\epsilon \to 0$. Quantities with a $\delta$ represent the first derivative with respect to $\epsilon$. To avoid divergences in Eq. (\ref{genvar}) associated with the logarithm, the null space of $\rho_A$ should be included in the null space of $\delta \rho_A$.}
\beq
\label{genvar}
\delta S_A = \delta \langle H_A \rangle \; .
\eeq
Here $H_A$ is the ``modular Hamiltonian" or ``entanglement Hamiltonian," defined to be the logarithm of the density matrix $\rho_A$ before the perturbation.

Applying this to an arbitrary CFT initially in its vacuum state and choosing the subsystem $A$ to be a ball $B$ of radius $R$, it is possible to find the modular Hamiltonian explicitly, as we review in appendix A. Using this result, (\ref{genvar}) becomes \cite{relative}
\beq \label{firstlawCFT}
\delta S_B = 2 \pi \int_{|x|<R} d^d x \frac{R^2 - |x|^2}{2R} \delta \langle T^{\text{CFT}}_{00}\rangle \equiv \delta E_B \; .
\eeq
Eq. (\ref{firstlawCFT}) is similar to the first law of thermodynamics in that it relates the change in entropy of a system to a change in some energy associated with the system.  However, in Eq. (\ref{firstlawCFT}), instead of the physical Hamiltonian of the global system, we find the modular Hamiltonian associated to the ball's density matrix. Further, the result (\ref{firstlawCFT}) makes no requirement that we are varying to a nearby equilibrium state. Thus, it is an exact quantum relation rather than a thermodynamic one.

\subsection{Entanglement in holographic CFTs}

Our goal is to understand the implications of the relation (\ref{firstlawCFT}) for the dual gravitational theory in the case of a holographic CFT. The ground state of the CFT is dual to an empty AdS spacetime with the bulk fields $\phi$ in their AdS ground state.  The appearance of AdS follows from symmetry considerations alone; we do not need to solve equations of motion to deduce the existence of this background. For some class of perturbations to the ground state, there will be a semi-classical dual spacetime characterized by some perturbed dual geometry and bulk state $(M, |\psi \rangle_{\text{bulk}})$.  The perturbed spacetime $M$ can be described by a metric in Fefferman-Graham form
\beq
ds^2 = \frac{\ell^2}{z^2}\left( dz^2 + \eta_{\mu \nu} dx^\mu dx^\nu + h_{\mu\nu} dx^\mu dx^\nu\right).
\eeq
We wish to derive constraints on the allowed $(h_{\mu\nu}, |\psi \rangle_{\text{bulk}})$ by translating the CFT constraint  (\ref{firstlawCFT}) to a relation in the dual theory using our basic assumption, that
\beq \label{holoee2}
S_A(|\psi\rangle_{\text{CFT}}) = \frac{|\tilde{A}|_M}{4 G_N} + S_\Sigma(|\psi\rangle_{\text{bulk}}),
\eeq

Before proceeding, we offer a few comments on this formula. The leading contribution $\frac{|\tilde{A}|_M}{4 G_N}$ to the entanglement entropy of the CFT is purely geometric, while the contribution due to bulk matter fields is suppressed in an expansion in $G_N$.  Indeed, $G_N \rightarrow 0$ is a classical limit, so matter fields in essence contribute the first quantum correction to the entanglement.  It has long been known in the context of black holes that the first quantum correction to the black hole entropy is obtained directly from the entanglement of field theory modes across the black hole horizon. Recently, FLM generalized this result to arbitrary minimal surfaces in the holographic context.  They argued that the first quantum correction to the RT prescription is essentially given by the bulk entanglement entropy of the bulk fields across the surface $\tilde{B}$.  However, we do not actually need the full power of the FLM analysis because the bulk region we consider, $\Sigma$, is very special.  Indeed, its boundary $\tilde{B}$ is a Killing horizon, so that the more traditional black hole arguments would suffice to tell us that the quantum correction is given by the bulk entanglement entropy across $\tilde{B}$. However, let us emphasize that we are simply assuming that Eq. (\ref{holoee}) is true.

In general, the various terms in (\ref{holoee2}) contain divergences that we now discuss.  The area law term has a single type of divergence associated with the fact the minimal surface extends all the way to $z=0$.  If we cutoff the geometry at $z=a$, with $a$ an effective microscopic length scale or UV cutoff in the CFT, then the resulting minimal surface formula has an $a$-dependence precisely in line with field theory expectations for the UV dependence of entanglement entropy. In this paper, we are concerned only with the variation of $S_{\text{CFT}}$ relative to the ground state. In this case, the corresponding variation of the bulk area is finite.

Turning now to the bulk entanglement term, we find now two types of divergences.  The first type of divergence comes again from the region near $z=0$ and may be dealt with in a similar fashion to the area law term.  We must, of course, specify boundary conditions of the fields at $z=a$, but this specification is part of the usual holographic dictionary.  The second type of divergence is a short distance bulk divergence.  This type of divergence has no obvious CFT interpretation and must be dealt with more carefully.  We expect that since the bulk is ultimately a quantum gravitational theory, the Planck length or some other short distance scale associated with gravity cuts off the bulk divergence in a natural way.\footnote{If we take the cutoff to be the Planck length then some terms in the so-called quantum correction are of the same order as the classical contribution.  However, these statements depend on the way in which we regulate the bulk entanglement entropy.  Also, we must include the bulk graviton, viewed as an ordinary QFT propagating in AdS, among the fields $\phi$.}  Cooperman and Luty, building on much earlier work, have recently shown that just such a program can be carried for all non-gravitational theories in perturbation theory \cite{Cooperman:2013iqr}.  More precisely, they showed that the structure of divergences in the bulk entanglement entropy was such that the gravitational sector provided precisely the right counterterms to absorb divergences.  Our needs are nevertheless somewhat different.  We will primarily be interested in variations of the bulk entanglement entropy, in which case one might expect that such divergences, which are a fixed property of the regulator, simply cancel as for the geometrical term.

\section{Argument for linearized equations coupled to matter}

In this section we present our main technical argument demonstrating that the linearized Einstein equations coupled to arbitrary matter fields emerge from the structure of entanglement in holographic CFTs.  We use the results and notation of \cite{Faulkner:2013ica}.  As discussed in the introduction, our starting assumptions about holographic entanglement are the Ryu-Takayanagi (RT) classical prescription \cite{rt1,rt2,timedep} and the Faulkner-Lewkowycz-Maldacena (FLM) \cite{Faulkner:2013ana} prescription for quantum corrections.

\subsection{Review of the leading order results}

In \cite{eom, Faulkner:2013ica}, it was shown using the leading order RT formula for holographic entanglement entropy that the general CFT result Eq. (\ref{firstlawCFT}) implies both that $\langle T_{\mu \nu}^{\text{CFT}}\rangle  \propto \lim_{z \to 0} z^{-d+1} h_{\mu \nu}(z)$ and that $h_{\mu \nu}(z,\vec{x},t)$ obeys the Einstein equations linearized about the AdS background. At this leading order in $G_N$, the bulk stress-energy tensor does not yet appear as a source for the metric.

\begin{figure}\label{setupfig}
\includegraphics[width=.2\textwidth]{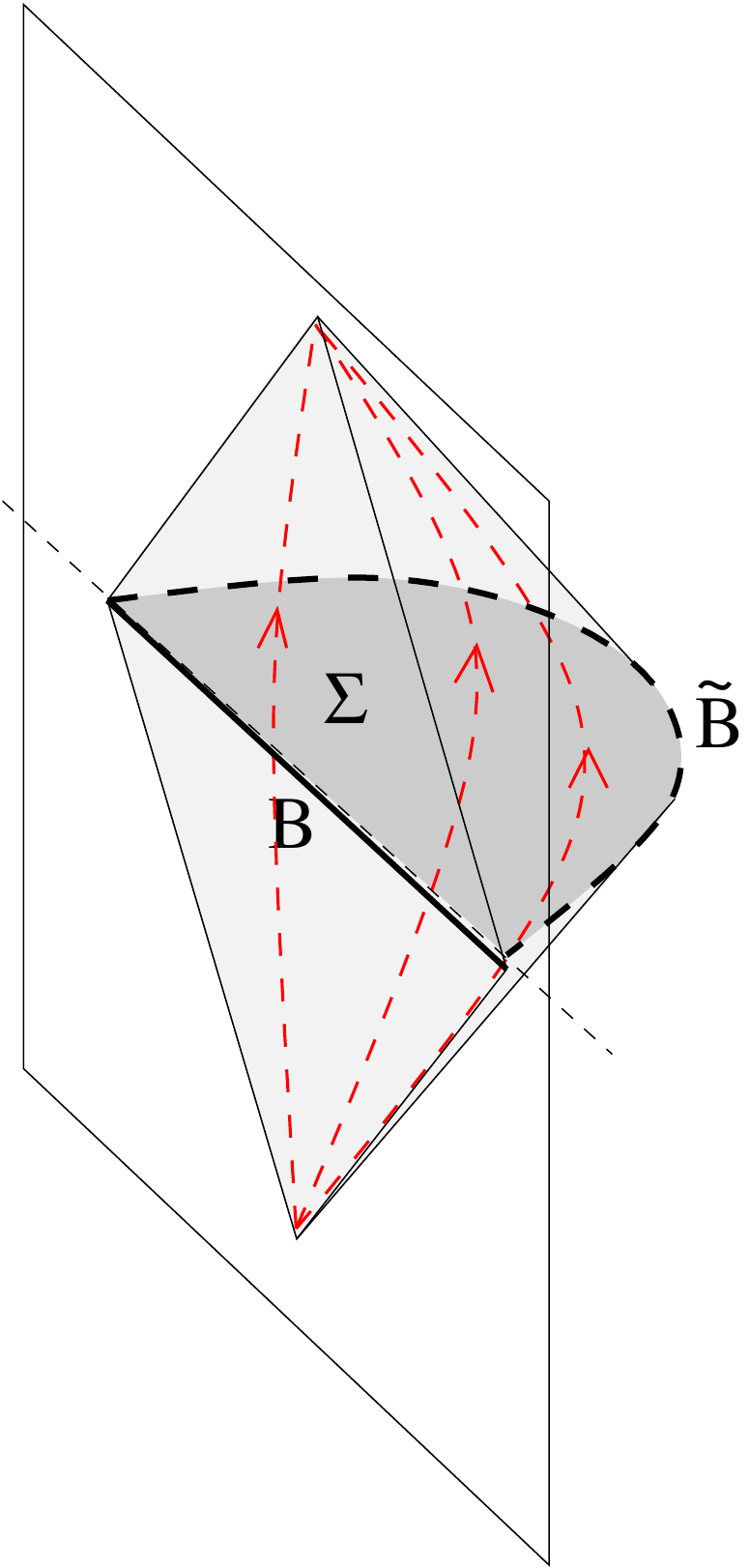}
\caption{Bulk setup for deriving local equations from the entanglement first law. Dashed curves show the flow generated by the Killing vector $\xi$.}
\end{figure}

For our generalization to the next order in $G_N$, we will take the same approach as \cite{Faulkner:2013ica}, so it will be useful to recall some of the detailed steps in the derivation. Let the boundary $d$-ball in the CFT be $B$, the bulk extremal surface be $\tilde{B}$, and the bulk region bounded by $\tilde{B}$ be $\Sigma_B$, as shown in figure 1.  The key step in the proof is the introduction of a $d$-form $\chi$ which satisfies
\beq
\label{chione}
\int_B \chi = \delta E_B^{\text{grav}},
\eeq
\beq
\label{chitwo}
\int_{\tilde{B}} \chi = \delta S_B^{\text{grav}},
\eeq
and
\beq
\label{chithree}
d \chi = - 2 \xi^a \delta E^g_{ab} \epsilon^b.
\eeq
In these formulae, $\delta E_B^{\text{grav}}$ and $\delta S_B^{\text{grav}}$ are the results of translating $\delta S_B$ and $\delta E_B$ to gravitational quantities using the holographic dictionary. They are integrals over $B$ and $\tilde{B}$ respectively of local quantities built from $h_{\mu \nu}$. The quantity $\delta E^g_{ab}$ is the linearized bulk Einstein equations with no bulk matter, $\epsilon^b$ is a volume form on $\Sigma$, and $\xi^a$ is a Killing field given by
\beq
\label{defxi}
\xi = - \frac{2\pi}{R} (t-t_0) [z \partial_z + (x^i - x_0^i) \partial_i ] + \frac{2\pi}{2R} [ R^2 - z^2 - (t-t_0)^2 - (x-x_0)^2] \partial_t .
\eeq
The flow generated by this Killing vector is shown in figure 1. We can easily check that $\xi$ is a Killing vector and that it vanishes on $\tilde{B}$, which is given by $R^2 =|\vec{x}|^2+z^2$ in the Fefferman-Graham coordinates.

Given these properties of $\chi$, the constraint $\delta E_B^{\text{grav}} = \delta S_B^{\text{grav}}$ implies that the integral of $\chi$ over the boundary of $\Sigma$ vanishes. By Stokes' theorem, this implies that $d \chi$ integrates to zero on $\Sigma$, and it follows by considering all possible balls $B$ in all Lorentz frames that $\delta E^g_{ab}$ is equal to zero at each bulk point \cite{Faulkner:2013ica}. This is the bulk linearized Einstein equations without matter.

\subsection{Bulk matter from $1/N$ corrections to holographic entanglement entropy.}

We now proceed to our main new result: that by taking into account the quantum correction term to the holographic formula as in (\ref{holoee}), the same basic CFT result (\ref{firstlawCFT}) implies that the stress energy tensor for the bulk fields appears as the source for the metric in the linearized gravitational equations.

\subsubsection{Corrections to $\delta S_B^{\text{grav}}$}

The bulk entanglement entropy contribution in (\ref{holoee}) introduces a new term into the bulk expression for the variation of entanglement entropy. This variation comes in principle from two sources: the geometrical data (extremal surface and metric) varies and the state of the bulk fields varies.  The variation of the bulk entropy with the minimal surface can be immediately dispensed with as it is zero to first order.  This follows because the bulk entanglement entropy, viewed as a function of the bulk surface, is also extremized on the same surface as the classical area law term.  This statement may be rendered plausible by noting that the simplest possibility for the bulk entanglement entropy is of course just the usual area law (identical to the classical geometrical form).  More generally, this follows from symmetry as we discuss in appendix C.  The variation with respect to the metric keeping the state fixed is also zero, since the entanglement entropy doesn't depend directly on the metric except possibly through counterterms (see appendix C).

The nonvanishing contribution to the variation of the bulk entanglement entropy is due to the variation of the bulk state $\delta |\psi\rangle_{\text{bulk}}$.  In general, the variation of the bulk entropy obeys the equation
\beq
\label{bulkvar}
\delta S_\Sigma(|\psi\rangle_{\text{bulk}}) = - \text{tr}( \delta \rho^{\text{bulk}}_{\Sigma} \log{(\rho^{\text{bulk}}_{\Sigma})}),
\eeq
where $\rho_\Sigma^{\text{bulk}} = \text{tr}_{\bar{\Sigma}}(|\psi\rangle\langle \psi |_{\text{bulk}} )$.  Now, the logarithm of the bulk density matrix (bulk modular Hamiltonian) is generically very hard to compute.  Remarkably, for the special choice of bulk region we are considering, which carries a natural geometric flow generated by the Killing field $\xi$, we can actually compute the bulk modular Hamiltonian. To understand this, we note that the region shown in figure 1 is a Rindler wedge of AdS. As a result, the modular Hamiltonian for $\Sigma$ can be computed by a parallel argument to the one showing that the half-space modular Hamiltonian for a QFT vacuum state on Minkowski space is the Rindler Hamiltonian (boost generator). We review this explicitly in appendix A.

Our result is that the modular Hamiltonian for $\Sigma$ is the AdS analogue of the Rindler Hamiltonian, which is precisely the generator of $\xi$. This can be written covariantly as\footnote{Note that on $\Sigma$, only $T_{00}$ contributes to this expression, since the only the 0 components of $\xi^a$ and $\epsilon^a$ are nonvanishing on this surface.}
\beq
\label{bulkdsde}
- \ln{(\rho^{\text{bulk}}_{\Sigma})} = \int_{\Sigma} \xi^a T^{\text{bulk}}_{ab} \epsilon^b \; ,
\eeq
so we have from (\ref{bulkvar}) that
\beq \label{eebulkvar}
\delta S_\Sigma(|\psi\rangle_{\text{bulk}}) = \int_{\Sigma} \xi^a \delta \langle T^{\text{bulk}}_{ab} \rangle \epsilon^b.
\eeq
Thus we have related the variation of the bulk entanglement entropy to the bulk stress tensor. We emphasize that all we needed to derive Eq. (\ref{eebulkvar}) was an expression for the modular Hamiltonian of the special region $\Sigma$ in the special geometry of pure AdS with the bulk fields in their vacuum state.

\subsubsection{Corrections to $\delta E_B^{\text{grav}}$}

We now argue that the gravitational expression for $\langle T^{\text{CFT}}\rangle$ is unchanged, so long as we restrict to perturbations $\delta \langle T^{\text{bulk}}_{ab} \rangle$ which vanish quickly near the boundary.

As discussed in \cite{Faulkner:2013ica}, it follows from (\ref{firstlawCFT}) that $\langle T^{\text{CFT}}(x)\rangle$ is proportional to the entanglement entropy $S_B$ in the limit of a small ball $B$ centered at $x$. Using our result (\ref{eebulkvar}) from the previous subsection, the correction to this entanglement entropy involves an integral of the bulk stress tensor over the region $\Sigma_B$ associated with this small ball. But this region $\Sigma$ becomes completely localized near the AdS boundary as the size of the ball goes to zero, and we have assumed that the variation of the bulk stress tensor vanishes here.

\subsection{The main argument}

We now have all the pieces in place to present the main argument.

For a general ball $B$, if we combine the leading order results (\ref{chione}), (\ref{chitwo}), and (\ref{chithree}) with the correction (\ref{eebulkvar}) to $S^{\text{grav}}$ and no correction for  $E^{\text{grav}}$ we obtain
\bea
\delta S^{\text{grav}} - \delta E^{\text{grav}} &=& \int_\Sigma \left( - 2 \xi^a \delta E^g_{ab} \epsilon^b + \xi^a \delta \langle T^{\text{bulk}}_{ab} \rangle \epsilon^b \right) \cr
&=& \int_\Sigma \left( - 2 \xi^a \left\{\delta E^g_{ab} - {1 \over 2} \delta \langle T^{\text{bulk}}_{ab} \rangle \right\} \epsilon^b \right).
\eea
Now, the CFT result (\ref{firstlawCFT}) implies the vanishing of the left hand side, and applying this to all balls following \cite{Faulkner:2013ica} it follows that
\beq
\label{result00}
- 2 \delta E^g_{00} + \delta \langle T^{\text{bulk}}_{00} \rangle = 0
\eeq
at each bulk point.

Recalling that $E_g$ was defined as
\beq
 E^g_{ab} = \frac{1}{\sqrt{-g}} \frac{\delta \mathcal{S}_g}{\delta g^{ab}}
\eeq
with
\beq
\mathcal{S}_g = \frac{1}{16 \pi G_N} \int d^{d+2} x \sqrt{-g} \left(R - 2 \Lambda \right),
\eeq
we see that (\ref{result00}) is the 00 component of the linearized Einstein equations with source term provided by the stress-energy tensor of the bulk quantum fields.

To obtain the other components of the Einstein equations in the field theory directions, we repeat the above analysis for various boosted frames labelled by velocity $u^\mu$.  We then conclude that $u^\mu u^\nu (- 2 \delta E^g_{\mu \nu} + \delta \langle T^{\text{bulk}}_{\mu \nu} \rangle )= 0$ for all $u^\mu$. To obtain the final $zz$ and $\mu z$ components, we follow \cite{Faulkner:2013ica} and appeal to an initial value formulation in a radial slicing.  Then these components represent constraints that if valid at $z=0$ are valid throughout the geometry.  They reduce at $z=0$ to demanding that the CFT stress tensor is conserved and traceless. See the appendix B for details.

It is useful here to clarify the meaning of the right hand side of the final result
\beq
\label{finalresult1}
\delta E^g_{ab} = {1 \over 2} \delta \langle T^{\text{bulk}}_{ab} \rangle \; .
\eeq
We first note that this formula sensibly involves the bulk stress tensor relative to the vacuum state, since presumably there is some background value which is sourcing the cosmological constant holding up the background AdS space. The stress-tensor appearing is the stress-energy tensor operator of the bulk quantum fields, including the metric perturbation.\footnote{If $\mathcal{S}_\phi$ is the bulk field theory action, then we may define the bulk stress tensor as
\beq
T^{\text{bulk}}_{ab} = -\frac{2}{\sqrt{-g}} \frac{\delta \mathcal{S}_\phi}{\delta g^{ab}}\bigg|_{g=g_{\text{AdS}}}.
\eeq
This object includes a graviton contribution because
\beq
\mathcal{S}_\phi \supset \frac{1}{2} \int \frac{\delta^2 \mathcal{S}_g}{\delta g^2}\bigg|_{g=g_{\text{AdS}}} h^2.
\eeq
Note that the graviton kinetic term is fixed by demanding that the action give the same equation of motion as the one obtained from the variation of (\ref{holoee}).} Since this typically starts with terms quadratic in the fields, we might worry that the right side (a first order variation) is identically zero. However, it is not difficult to see that perturbations to the quantum state will generically give a nonzero result for the right side. The quadratic part of the local stress tensor operator, expanded in terms of creation and annihilation operators for the bulk fields includes terms of the form $a_{k_1} a_{k_2}$ and $a_{k_1}^\dagger a_{k_2}^\dagger$ in addition to $a_{k_1}^\dagger a_{k_2}$ terms. A generic variation of the bulk state will include terms such as $|\psi\rangle_{\text{bulk}} = |0 \rangle + \lambda c_{k_1 \; k_2} a_{k_1}^\dagger a_{k_2}^\dagger |0 \rangle + \dots$ where $\lambda$ is the small parameter defining the infinitesimal variation. In this case, the expression $\langle T^{\text{bulk}}_{ab} \rangle$ will receive contributions at first order in $\lambda$, so that $\delta \langle T^{\text{bulk}}_{ab} \rangle$ is nonvanishing.

On the other hand, classical perturbations to the bulk fields (e.g. a classical scalar field profile with amplitude $\lambda$) contribute to $\langle T^{\text{bulk}}_{ab} \rangle$ only at second order in $\lambda$. Therefore, these are not directly captured by the right side of (\ref{finalresult1}). However, the fact that $\delta \langle T^{\text{bulk}}_{ab} \rangle$ appears in the equation governing linearized perturbations suggests strongly that $\langle T^{\text{bulk}}_{ab} \rangle$ appears more generally as a source for the gravitational field. Specifically, we can say that if the source term were some other operator $\langle T^{\text{bulk}}_{ab} + {\cal O}^{\text{bulk}}_{ab}\rangle$, the linear variation $\delta \langle {\cal O}^{\text{bulk}}_{ab}\rangle$ must vanish for arbitrary perturbed bulk states corresponding to our class of CFT perturbations; otherwise, there would be additional contributions to (\ref{finalresult1}). This suggests that the operator ${\cal O}^{\text{bulk}}_{ab}$ must annihilate the vacuum state when acting from the left or right. Within the class of local quantum field theory operators, there is no such possibility (e.g. since we always have terms with creation but no annihilation operators).

To summarize, the direct result of our analysis is that the linearized source term in the bulk gravitational equation is $\delta \langle T^{\text{bulk}}_{ab} \rangle$. With the additional assumption that the source is a local quantum field theory operator, we have argued that it can only be $T^{\text{bulk}}_{ab}$.

\section{Discussion}

Starting from the holographic entanglement entropy formula (\ref{holoee2}), we have shown using CFT relation (\ref{firstlawCFT}) that for any perturbation to the CFT vacuum state with a semi-classical gravitational dual description $(M,|\psi\rangle_{\text{bulk}})$, the metric perturbation describing $M$ is constrained by the linearized Einstein equation including the source term,
\beq
\label{finalresult}
\delta E^g_{ab} = {1 \over 2} \delta \langle T^{\text{bulk}}_{ab} \rangle \; .
\eeq
In this final section, we offer several comments on the result.

\subsection{Domain of validity of the result}

An unsettling feature of the final formula (\ref{finalresult}) is that the right hand side includes a quantum expectation value while the left side is classical. Since our dual gravitational theory is supposed to be a complete theory of quantum gravity, it might be expected that Einstein's equations should hold as an operator relation. For example, if we consider a quantum state that is the superposition of states in which a star is at one location and a state in which a star is at another distant location, the spacetime geometry should also exist as a quantum superposition of metrics rather than a single metric sourced by the quantum expectation value of the matter. On the other hand, the result (\ref{finalresult}) suggests the latter unphysical possibility.

The resolution of this apparent difficulty is that we have assumed in our derivation that the CFT states under consideration have dual gravitational descriptions which are semi-classical. That is, we have assumed that the dual spacetime is well-described by a single classical geometry with quantum fields living on it. This restriction was necessary because the holographic entanglement entropy formula (\ref{holoee}) applies only with such a restriction; even the leading order Ryu-Takayanagi formula is ill-defined for a state in which the bulk state is a quantum superposition of two different geometries.

Thus, the result (\ref{finalresult}) should be understood as a constraint on states of the dual gravitational theory that have a good semiclassical description.

\subsection{Origin of the universality of gravity}

Our final result includes the familiar statement that all sources of stress-energy act as a source for gravity. It is interesting to pinpoint the origin of this fundamental result, since we have now derived it from more basic principles in the context of holographic CFTs. The stress tensor appeared in our derivation when we wrote down an explicit expression for the bulk entanglement across the surface $\tilde{B}$. The reason that all sources of stress-energy appeared is that all bulk degrees of freedom contribute to this bulk entanglement. This is automatic, since entanglement by its definition is an inclusive quantity that takes into account all degrees of freedom in the subsystem under consideration. Thus, we can say that {\it the universality of the gravitational interaction comes directly from the universality of entanglement} - it is not possible to have stress-energy that doesn't source the gravitational field because it is not possible to have degrees of freedom that don't contribute to entanglement entropy.

\subsection{From the linearized equations to the non-linear equations}

The coupling to matter in the linearized equation (\ref{finalresult}) can be cast in the Lagrangian formulation as $ \mathcal{L}_{\text{coupling}} = - \frac{1}{2} T_{ab}^{\text{bulk}} h^{ab}.$ The presence of the gravitational field $h$ modifies the equations of motion of matter via this coupling. In particular, we have the phenomenon of energy transfer from the matter degrees of freedom to the linearized spin two field. Furthermore, since quadratic terms in $h$ are included in $T^{\text{bulk}}$, the modified action necessarily also contains graviton self-interactions.  To preserve bulk energy conservation we are immediately forced into a non-linear or self-coupled version of the theory. It is well known \cite{Feynman,Deser:1969wk,Deser:1987uk,Wald:1986bj} that in many situations, the self-consistent endpoint of this analysis is the full non-linear Einstein's equations, or some other generally covariant theory (e.g. with higher powers of curvature). In our case, Einstein's equations would be singled out as the unique choice given our starting assumption that the leading contribution to holographic entanglement entropy is area. The holographic entanglement functional applied to a black hole horizon must give the black hole entropy, and in the more complicated examples, black hole entropy is not simply the horizon area but some other Wald functional computed from the gravitational action \cite{wald0,iyer1}. As shown in \cite{Faulkner:2013ica}, had we started with a more general functional we would have obtained the linearized equations of the corresponding higher curvature theory. In this case, the covariant action we would obtain in the end would be the corresponding covariant higher curvature theory.

A major caveat here is that the arguments leading from linearized gravity coupled to matter to the full Einstein equations make the assumption that the gravitational theory continues to be described by some local Lagrangian in the nonlinear regime. This is certainly plausible in our case since we have shown this directly for the linearized theory. However, we currently do not have an argument directly from the CFT that it has to be true away from the linear regime. We leave the question of deriving bulk locality for future work.\footnote{Specifying the rules of the game for considering non-local bulk theories and trying to restrict the allowed possibilities in such a more general setting is interesting, but we do not pursue it here. We also note that a ``non-local" bulk theory could mean at least two different things.  First, it could mean that there are explicitly non-local terms in the action.  Second, it could mean that the action is local but the number of bulk fields is sufficiently large, etc. that \textit{physics} is effectively non-local.  It seems that only the first more extreme kind of non-locality is incompatible with the statement that we have a ``local Lagrangian".}

\subsection{Holographic entanglement dictionary from entanglement renormalization}

Let us now return to our primary assumption, namely that for a class of holographic CFTs, the entanglement entropy of a ball $B$ in the field theory is the sum of two pieces,
\beq
S_B(|\psi\rangle_{\text{CFT}}) = \frac{|\tilde{B}|_M}{4 G_N} + S_\Sigma(|\psi\rangle_{\text{bulk}}),
\eeq
where $S_\Sigma(|\psi\rangle_{\text{bulk}})$ is the entanglement entropy of bulk fields across the surface $\tilde{B}$.

The validity of this formula has been argued assuming the AdS/CFT correspondence in \cite{aitor,Faulkner:2013ana}. However, it is interesting to ask whether such a formula can be derived or at least motivated directly from the CFT. Is it possible to show directly that for certain CFTs, the entanglement structure of low-energy states can be consistently captured by dual spacetime configurations via such a relation?

A promising step in this direction is the idea of entanglement renormalization \cite{Vidal:2007hda}.  This is a general method for representing the ground state (and more general states) in terms of a hierarchical entanglement structure known as MERA (multiscale entanglement renormalization ansatz) which is a special case of a so-called ``tensor network".  It has already been shown that the resulting network structure behaves like a discrete graph analog of (the spatial slice of) AdS when the system is conformal \cite{swingle}.  It has also been shown that an ``area law" in the discrete AdS network bounds the amount of entanglement any region can have with its complement.  Furthermore, in simple numerical studies it has been observed that this bound is actually saturated.  Finally, early steps have been taken to argue in the large N and strongly coupled limit, the bulk area law bound is indeed saturated \cite{swingle}.  Along with a demonstration that the bulk geometry is effectively smooth and local, this would constitute a demonstration of the holographic area law, the first term in Eq. (\ref{holoee}).

We have already commented in the introduction that once we have the bulk geometry and the area formula, the addition of the bulk matter term across the horizon is quite natural.  However, it should also be possible to derive this term directly from CFT considerations.  Recently, some progress has been made in this direction by considering explicit bulk degrees of freedom in the MERA setup.  In fact, we can show within the MERA construction that if the so-called bulk degrees of freedom are additionally entangled on top of the basic network, then the entropy of the resulting CFT state decomposes into the sum of two terms, an ``area" term from the network and a bulk entanglement term.  More precisely, the bulk entanglement term is the ``long-range" part of the bulk entanglement entropy and does not include short distance entanglement within the bulk of the entanglement network.  These developments will be discussed in detail in a forthcoming paper.

Optimistically, we can envision the following outcome.  Given a semi-classical CFT state $|\psi\rangle_{\text{CFT}}$ (such as the ground state) in, say, a regulated lattice model, we can in principle construct an entanglement network to represent the state.  The entanglement network has the property that the entanglement entropy of a region in the boundary is given by an ``area" term plus a bulk entanglement term quite analogous to Eq. (\ref{holoee}). If we can complete the technical challenge of showing that for certain special CFTs the entanglement network really limits to a smooth and local geometry, and if we can further show that the ``area" term is precisely the area of a minimal surface in this geometry, then using the arguments in this work we would have a complete construction of dual gravitational equations from purely CFT considerations.

\section{Acknowlegements}

We thank Thomas Hartman and Rob Myers for valuable comments on a preliminary version of the draft and John McGreevy for helpful discussions. BGS is supported by a Simons Fellowship through Harvard University. MVR is supported in part by the Natural Sciences and Engineering Research Council of Canada and by FQXi. MVR would like to thank Harvard University for hospitality during the early stages of this work.  BGS would like to thank the Simons Institute for the Theory of Computing at Berkeley for hospitality while this work was completed.

\appendix

\section{The entanglement ``first law.''}

In this appendix, we begin by recalling the derivation of the entanglement ``first law,'' including a slight generalization that we require in the main argument. For a general quantum system, varying the global state in the definition $S_A = - \text{tr}(\rho_A \log (\rho_A))$ of the entanglement entropy for a subsystem $A$ gives the general relation
\beq
\label{dSdK}
\delta S_A = \delta \langle H_A \rangle \; ,
\eeq
where $H_A = -\log(\rho_A)$, the ``modular Hamiltonian'' or ``entanglement Hamiltonian,'' is defined as the logarithm of the unperturbed density matrix.

Now, consider a general quantum field theory on a spacetime M, in a state defined using a particular Euclidean action $S_E$ via
\beq
\langle \phi_0 |\Psi \rangle = \int_{\phi(0) = \phi_0}  [d \phi(\tau < 0)] e^{-S_E} \; ,
\eeq
where the integral is over fields $\phi$ on the Euclidean continuation $M_E$ of the geometry $M$. For example, this defines the vacuum state if we choose $M$ as Minkowski space or AdS. The density matrix for some region $A$ on the $\tau=0$ slice is given by
\beq
\label{dm1}
\langle \phi_1 |\rho_A| \phi_0\rangle = \int^{\phi(0+) = \phi_0}_{\phi(0-) = \phi_1}  [d \phi(x \not \in A)] e^{-S_E} \; .
\eeq
Now, suppose that in the Euclidean geometry $M_E$ we have a $U(1)$ isometry parameterized by an angular coordinate $\theta$, such that the boundary of $A$ is a fixed point for the isometry and the Euclidean action $S_E$ is also invariant. In the case of a conformal field theory, we can also consider the case where $\theta$ parameterizes a conformal isometry direction. Then we can reinterpret $\theta$ as a Euclidean time coordinate, and rewrite the right hand side of (\ref{dm1}) as
\beq
\int^{\phi(\theta =0) = \phi_0}_{\phi(\theta = 2 \pi) = \phi_1}  [d \phi(\theta)] e^{-S_{\theta}} \; .
\eeq
where $S_{\theta}$ is the Euclidean action associated with the generator $H_\eta$ of translations in the associated Lorentzian coordinate $\eta = i \theta$.

We recognize this as the path-integral expression for the matrix elements of a thermal density matrix $e^{-\beta H_\eta}$ where $\beta = 2 \pi$.

Thus, for spacetimes M and regions $A$ to which our conditions apply, we have
\beq
\rho_A = e^{-H_A} = e^{-2 \pi H_\eta} \; ,
\eeq
so from the general result (\ref{dSdK}), we obtain
\beq
\delta S_A = 2 \pi \delta \langle H_\eta \rangle \; .
\eeq
To obtain the result (\ref{bulkdsde}), we choose $M$ as pure anti de Sitter space, and $A$ as the region $\Sigma$. We need only verify that after analytic continuation, the Killing field (\ref{defxi}) generates a $U(1)$ isometry of the Euclidean AdS with $\partial \Sigma$ fixed. This is simplest to see if we choose Poincar\'e coordinates $ds^2 = (dz^2 + d\tau^2 + dx_i^2)/z^2$ for the Euclidean continuation of AdS such that $\Sigma$ is described by $\{\tau = 0,x_1>0\}$. In these coordinates, the flow associated with the Euclidean continuation of $\xi$ is simply the rotations about the origin in the $x_1-\tau$ plane.

\section{Constraint equations}

In this appendix, we show that the $z \mu$ and $zz$ components of Einstein's equations coupled to matter will be satisfied throughout the bulk provided that they are satisfied at $z=0$ and that the $\mu \nu$ equations are satisfied everywhere. We work at the non-linear level but all equations can be linearized if desired.

Defining $W_{ab} = R_{ab} - {1 \over 2} g_{ab} R + \Lambda g_{ab}$ to be the Einstein tensor plus cosmological constant term, our assumptions are that
\[
W_{\mu \nu} = C T_{\mu \nu}
\]
and that the bulk stress tensor is conserved
\[
\nabla^a T_{ab} = 0 \; .
\]
Now, consider the other components $W_{z a} - C T_{z a}$. We will show that if these vanish at $z=0$, they vanish everywhere. We have
\bea
\nabla_z (W_{z a} - C T_{z a}) &=& \nabla_z W_{z a} - C \nabla_z T_{z a}) \cr
&=& -\nabla^\mu W_{\mu a} + C \nabla^\mu T_{\mu a} \
\eea
where we have used the contracted Bianchi identity
\[
\nabla^a W_{ab} = 0 \; .
\]
For $a = \nu$, this gives
\bea
\nabla_z (W_{z \nu} - C T_{z \nu}) &=&  -\nabla^\mu W_{\mu \nu} + C \nabla^\mu T_{\mu \nu} \cr
&=& -C \nabla^\mu T_{\mu \nu} + C \nabla^\mu T_{\mu \nu} \cr
&=& 0
\eea
where we have used the $\mu \nu$ equation. Thus, we conclude that
\[
W_{z \nu} = C T_{z \nu}
\]
holds everywhere if it holds at $z=0$ (which is guaranteed by conservation and tracelessness of the boundary stress tensor).
For $A = z$, the equation above gives
\bea
\nabla_z (W_{z z} - C T_{z z}) &=&  -\nabla^\mu W_{\mu z} + C \nabla^\mu T_{\mu z} \cr
&=& -C \nabla^\mu T_{\mu z} + C \nabla^\mu T_{\mu z} \cr
&=& 0
\eea
where we have used the recently derived $z \nu$ equation. So
\[
W_{z z} = C T_{z z}
\]
holds everywhere if it holds at $z=0$ (again guaranteed by conservation and tracelessness of the boundary stress tensor).

\section{Properties of bulk entanglement}

In this appendix, we provide more detailed arguments for two assertions made in the main text.

\subsection{Bulk entanglement entropy is extremized on horizon}

Let us consider the bulk entanglement entropy $S_\Sigma(|\psi\rangle_{\text{bulk}})$ as a function of the bulk region $\Sigma$ with the stipulation that $\Sigma$ asymptotes to $A$ at the boundary of AdS.  Then we claim that $S_\Sigma(|\psi\rangle_{\text{bulk}})$ is extremized when $\Sigma$ is bounded by the extremal surface $\tilde{A}$ which is also the Killing horizon.  This follows from symmetry since the bifurcation surface $\tilde{A}$ essentially divides the space in half.  In Fig. 2 we see in panel (1) the geometrical situation of interest.  In panel (2) we see the result of a conformal transformation of the situation in panel (1).  In the situation in panel (2) it is clear from symmetry that the red surface (analog of $\tilde{A}$) will extremize the bulk entanglement entropy.

\begin{figure}\label{eeconfmaps}
\includegraphics[width=.6\textwidth]{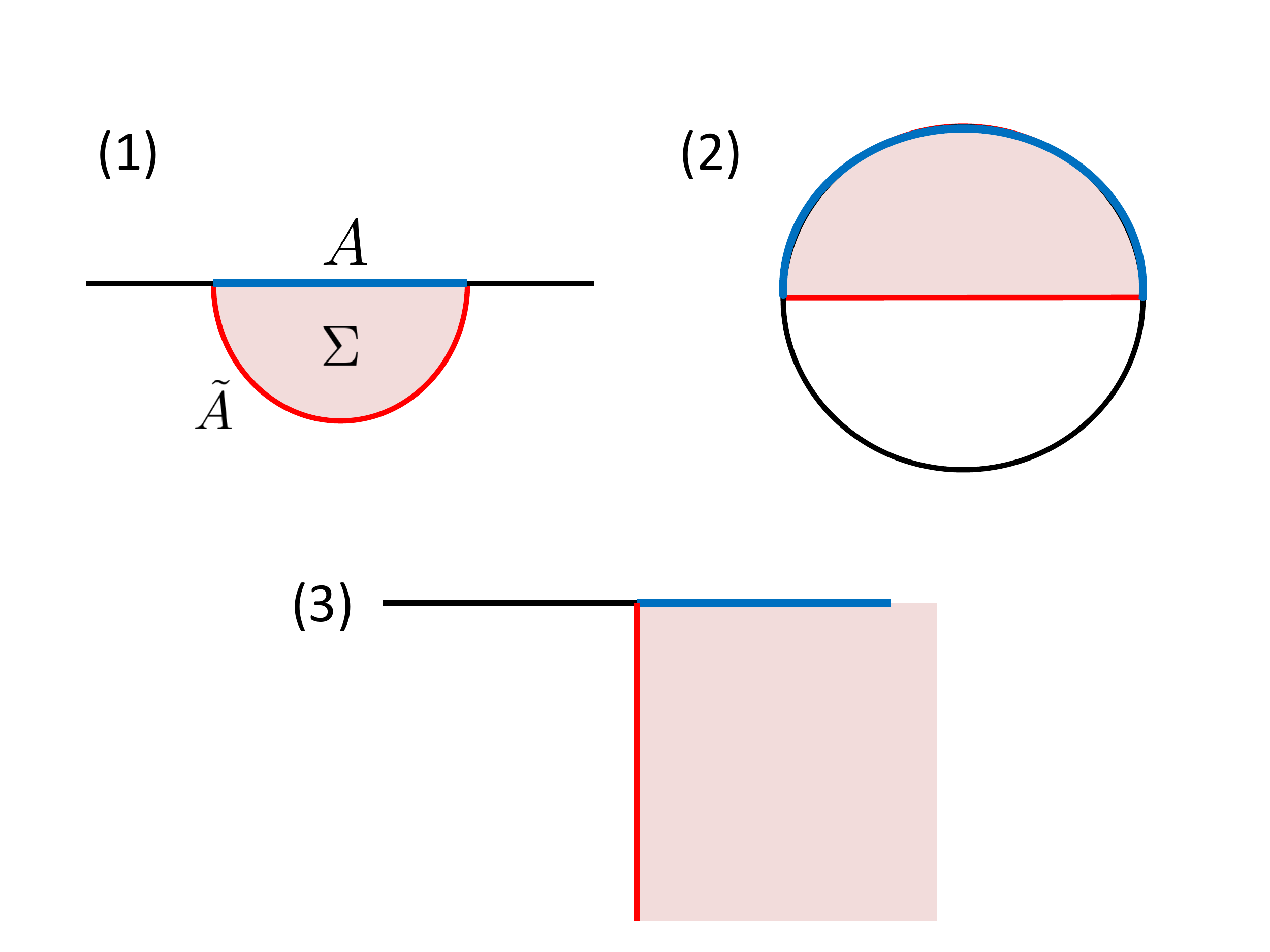}
\caption{Panel (1) shows the geometrical setup we are considering.  Panel (2) represents the effect of a conformal transformation (whose inverse is analogous to stereographic projection) which maps the infinite line to a sphere and the region $A$ to half the sphere.  Panel (3) represents the effect of yet another conformal transformation where this time we send one of the intersections of $A$ and $\tilde{A}$ to infinity.}
\end{figure}

\subsection{Bulk entanglement entropy depends only implicitly on bulk metric}

Here we will show that the bulk entanglement entropy does not depend directly on the metric, even when it is properly regulated.  The basic argument is very simple.  Fixing a system of coordinates, if we keep the entangling region $A$ fixed and the state of the fields fixed, the metric is completely irrelevant.  In other words, the entanglement only depends indirectly on the metric through the state.  Said differently still, if we add an operator to the Hamiltonian that corresponds to a change in the metric, but we don't change the actual physical state or the region of interest, then the entanglement is unchanged.

\end{document}